\documentclass[10pt,final,conference]{IEEEtran}

\ifCLASSINFOpdf
\else
\fi
\usepackage{graphicx}
\usepackage{epstopdf}
\DeclareGraphicsExtensions{.pdf,.jpeg,.png}
\DeclareGraphicsExtensions{.eps}
\newcounter{example}[section]

\graphicspath{{Figures/}}
\usepackage[super]{nth}
\usepackage{romannum}

\usepackage{amssymb}
\usepackage{verbatim}
\usepackage{bbm}

\usepackage[cmex10]{amsmath}
\usepackage{amssymb}
\usepackage{algorithm,algcompatible}
\usepackage{lipsum}
\usepackage{algpseudocode}
\usepackage{algorithmicx}
\usepackage{array}
\usepackage[font=footnotesize]{subfig}
\usepackage{stfloats}
\usepackage[font=footnotesize]{caption}
\usepackage{amsmath}
\usepackage{amsthm}
\usepackage{multirow}
\usepackage[table,xcdraw]{xcolor}

\usepackage{breqn}
\usepackage{tabularx}
\usepackage{tabu}
\usepackage{tabulary}
\usepackage{amsmath}
\usepackage{multicol}

\usepackage[english]{babel}

\algnewcommand\INPUT{\item[\textbf{Input:}]}%
\algnewcommand\OUTPUT{\item[\textbf{Output:}]}%

\begin{document}
\title{Online Shaping for ISI Channels with a Limited Number of ADC Bits}
\author{\IEEEauthorblockN{Or Levi}
\IEEEauthorblockA{Dept. of EE-Systems, TAU\\
Tel-Aviv, Israel\\
Email: orl1@mail.tau.ac.il}
\and
\IEEEauthorblockN{Dan Raphaeli}
\IEEEauthorblockA{Dept. of EE-Systems, TAU\\
Tel-Aviv, Israel\\
Email: danr@eng.tau.ac.il}}

\maketitle

\begin{abstract}

%An online shaping technique for high performance communication over Gaussian channels with intersymbol interference (ISI) and receiver analog to digital converter (ADC) quantization noise is presented. The suggested technique uses precoding, over a Pulse Amplitude Modulation (PAM) constellation at the transmitter, such that symbol sequences which would cause high peak power values at the receiver are avoided. The results are a reduction in the peak to average power ratio (PAPR) at the ADC input and, interestingly, a reduction in the ISI level at the receiver. 
%On a practical scenario, aim to transmit 200 Gbps on a 50 cm Printed Circuit Board (PCB) trace, we demonstrate in simulations an overall shaping gain as high as 8.55 dB, compared to uniform transmission with turbo equalization at the receiver side. 

An online shaping technique for high performance communication over Gaussian channels with Inter-Symbol Interference (ISI) and receiver Analog to Digital Converter (ADC) noise is presented. 
The technique uses online transmitter precoding over Pulse Amplitude Modulation (PAM) constellation, designed to shape the symbols distribution so that peak power constraint at the channel output is satisfied. An iterative decoder shares information between a modified M-BCJR module, which computes online the trellis transition probabilities of the shaped distribution, and turbo decoder. The result is a reduction in the required ADC Effective Number Of Bits (ENOB), which is in particular attractive in modern high-speed wireline links. 
Theoretical bounds are analytically derived which enable to assess the possible gain using shaping.
On practical scenarios aim to transmit 200 Gbps and 400 Gbps over printed circuit board, we demonstrate in simulations an overall ENOB gains as high as 1.43 bit and 1.78 bit, respectively, compared to uniform 4-PAM transmission with turbo equalization at the receiver side.

\end{abstract}

%===============================================================================
\section{Introduction}
\label{intro}
%===============================================================================%\IEEEPARstart{W}ireline communication channels such as, chip-to-chip SerDes communication channels \cite{SerDes1}, optical communication channels, and magnetic recording channels, can be modeled by Additive White Gaussian Noise (AWGN) channels with Inter-Symbol Interference (ISI) \cite{FroneyISI}. Recently, advanced very high-speed links are starting to employ ADCs at the receiver to convert the analog signal to a digital signal for using advanced digital receiver \cite{adc_mot1}, \cite{adc_mot2}.

 %\IEEEPARstart{R}ecently, advanced very high-speed communication links, such as chip-to-chip SerDes communication channels \cite{SerDes1}-\cite{SerDes3} and optical communication channels, have started to employ ADCs at the receiver to convert the analog signal to a digital signal for using advanced digital receiver \cite{adc_mot1}-\cite{adc_mot3}. 
%The ever increasing demand for higher data rates in such channels is not going to stop in the foreseen future.
%Raw data rates of 100Gbps, 200Gbps and 400Gbps are intended to be used in the next Ethernet generation \cite{HighData1}. 

\IEEEPARstart{T}he ever increasing demand for higher data rates in wireline communication links imposes the use in sophisticated digital equalization techniques, usually implemented at the receiver side \cite{adc_mot3}. Those, require using high-speed front-end ADCs for proper analog to digital signal conversion.   %implemented in back-plane receivers that rely on a  followed by a digital equalizer \cite{adc_mot3}.
%\IEEEPARstart{T}he ever increasing demand for higher data rates in wireline communication links, such as chip-to-chip SerDes communication channels \cite{SerDes1} and optical communication channels, imposes the use in sophisticated digital equalization techniques, implemented in back-plane receivers that rely on a front-end ADC followed by a digital equalizer \cite{adc_mot3}.
%It is well-known that a major non-ideal issue of wireline links is the frequency dependent channel loss. Such a behaviour causes an increment in the ISI level at the channel output as data rate increases which in turn causes an increment in the PAPR of the received signal since, according to Central Limit Theorem (CLT), the symbols distribution at the channel output approaches the Gaussian distribution.

It is well-known that a major non-ideal issue of wireline links is the frequency dependent channel loss. Such a behaviour causes an increment in both ISI and PAPR of the signal at the channel output as baud rate increases.
% This increment could be explained by the Central Limit Theorem (CLT).
% as the signal distribution approaches the Gaussian distribution.
%For example, a comparison between PAPR distributions at the channel output in different transmission rates is shown in Fig. \ref{PAPRU}. In this example, the transmission is over SerDes channel (typical 50cm chip to chip trace), and the transmitted symbols are chosen uniformly from a 4-PAM constellation. It can be seen that the PAPR in these cases, depending on the transmission rate, may reach well over 10 dB. This channel, with symbol rate 112 Gsymbol/sec, is denoted Channel-A.
%\begin{figure}[ht]
%\centering
%\includegraphics[width=2.3in]{PAPR_uniform_4_pam_diff_rates}
%\caption{PAPR distributions at the channel output in a different symbol rates.}
%\label{PAPRU}
%\end{figure}
%However, the channel impulse response spans over higher symbol periods as the symbol rate gets higher on the same physical channel. Hence, according to the central limit theorem, the symbols distribution at the channel output approaches the Gaussian distribution, which is characterized by its very high PAPR.   
To avoid excessive signal distortion due to clipping the ADC is required to supply large dynamic range, which leads to high ENOB \cite{IEEEADC} requirement to achieve the desired system performance. The demand of large dynamic range translates to higher circuit design complexity and higher power consumption, which are among the key issues in high-speed applications.

Fischer proposed Dynamics Limited Precoding (DLP) technique \cite{thp_ext} that allows receiver PAPR control.
%In \cite{thp_ext}, a method called Dynamics Limited Precoder (DLP) to control the PAPR at the receiver was introduced. 
DLP is an extension of the well-known Tomlinson-Harashima Precoder (THP) \cite{thp1}, which offers a trade-off between transmitter and receiver PAPR. One extreme point of DLP is original THP, with minimal PAPR at channel input and maximal PAPR at channel output. The other extreme point is essentially channel inversion at the transmitter which provides minimal PAPR at channel output in expense of maximal PAPR at channel input. Since channel input (transmitter output) is voltage limited and quantization prone due to the Digital to Analog Converter (DAC), DLP shifts the problem to the transmitter side without providing any gain overall. On the contrary, the quantization noise at the transmitter is an additional noise source.

A common equalizer that can be used for ISI (and PAPR) reduction at the ADC input is a Continuous Time Linear Equalizer (CTLE) \cite{ctler}. CTLE has several disadvantages. First, CTLE introduces large impedance discontinuity at the channel and equalizer interface. Impedance matching networks, often employ inductors, can be used to prevent the discontinuity. However, the large inductors make this approach less suitable for on-chip integration. %Second, since equalization with passive CTLE is performed by attenuating low-frequency signal spectrum, the average power of the received signal is reduced. It is desirable therefore to have a gain greater than one at some frequencies to maximize the benefit from receiver-side equalization. Hence, CTLE using active circuit elements rather than passive components are required. However, active CTLE amplifies the high frequency noise which potentially degrading the receiver's noise figure. 
In addition, CTLE must be optimized for each channel and both devising adaptation algorithm and practically modifying the components at high frequencies are formidable challenges. %\cite{ctle2}.%, \cite{ctle2}.

Inspired by the mathematical similarity between the problem at hand and the problem of PAPR reduction at the transmitter due to the pulse shaping filter effect, we sought to derive a parallel technique. 
%In \cite{papr2}, a shaping scheme for PAPR reduction at a pulse shaping filter output was presented.
A recent shaping technique for PAPR reduction at the transmitter was presented in \cite{papr2}.
To avoid peak excursions at the pulse shaping filter output, symbol transitions which result in high peak values are removed from the trellis graph, so that PAPR gain is achieved compared to un-shaped transmission. However, both implementation and theoretical analysis require a prior calculation of the shaped distribution which is stored in a table. The table size depends exponentially on the pulse shape filter span. Hence, it cannot be used for practical long channels due to the enormous size of the required memory.
%In this paper we essentially expand this approach by deriving theoretical bounds, which enables to asses the possible gain using shaping, and by deriving practical online shaping scheme. Both can be used for practical long channels.
%The online shaping scheme, similar to \cite{papr2}, is based on modeling the transmission over the channel as a Markov process.
%Whereas \cite{papr2} requires a prior calculation of the shaped distribution and storage in a table with a size that exponentially depends on the channel length, the proposed shaping scheme is designed to eliminate this demand by employing online calculation of the shaped distribution (both at transmitter and receiver). 

In this paper, we propose an online shaping scheme for PAPR reduction at the output of wireline channels that enables to reduce the ADC ENOB requirement compared to un-shaped transmission. By theoretical analysis we derive an upper-bound on the shaping gain and we show that the proposed scheme approaches it. 
The shaping scheme is attractive especially in high-speed links due to high PAPR at the ADC input on the one hand, but limited receiver power consumption and complexity on the other hand.
Whereas \cite{papr2} requires high memory and a prior calculation of the shaped distribution, the proposed shaping scheme is designed to eliminate these demands by employing online calculation of the distribution (both at transmitter and receiver). It can therefore be used for practical long channels. 
%The proposed shaping scheme employs online calculation of the shaped distribution (both at transmitter and receiver). Therefore can be use for practical long channels. 
%At the transmitter, the precoder examines all the constellation symbols, considering the past symbols that already have been transmitted, and selects a symbol that satisfies the peak power constraint at the channel output. As a result, the PAPR and the ISI at the channel output are reduced. 
%At the receiver side, we use a reduced complexity M-BCJR module. The M-BCJR is a BCJR algorithm with only $M$ best survivors kept at each forward stage \cite{MBCJR}. 
%Since the distribution that satisfies the peak constraint at the channel output is calculated online then, as will be shown in the sequel, the proposed online shaping scheme could be used for practical long channels (and thus, in very high data rates). 
%Additionally, we will show that the shaping scheme shapes the transmitted spectrum as high pass, without using any filter, with a slope that depends on the peak constraint. At low enough peak constraint, the process leads to a transmitted spectrum
%which approximates the channel inverse spectrum, so that the receiver spectrum becomes almost flat
Since the suggested shaping scheme uses transmitter precoding over a standard PAM constellation, and does not use any filter, then, unlike transmitter equalization and DLP it does not increase either transmitter PAPR or number of signal levels at the transmitter. Therefore, it has no effect on the required transmitter hardware (e.g., DAC and power amplifier).
For data rates 200 Gbps and 400 Gbps, a shaped 8-PAM transmission achieves ADC ENOB gains of 1.43 bit and 1.78 bit, respectively, compared to uniform 4-PAM transmission with Turbo-Equalizer (TE) \cite{TE} at the receiver. 

The rest of the paper is organized as follows: Section \ref{sm} describes the system model. In Section \ref{Implementation} we present the online shaping process, both for the Tx and Rx parts. In Section \ref{UB} we present a theoretical analysis and derive the achievable gains using shaping. Section \ref{Results} presents simulation results of the shaped system, compared to a uniform transmission. In Section \ref{conclusion} conclusion remarks are given.

%Examples of uniform and shaped 8-PAM distributions at typical 50cm chip to chip trace output are illustrated in Fig. \ref{shaped_unshaped_PDF}. 

%\begin{figure} [ht] 
%\centering
%\subfloat[]{{\includegraphics[width=3.15in]{un_shaped_D_200.eps} }}%
%\subfloat[]{{\includegraphics[width=3.15in]{shaped_peak_0055_D_200.eps} }}%
%\qquad
%\caption{ Distribution at channel output (a) Uniform transmission. (b) Shaped transmission.}
%\label{shaped_unshaped_PDF}
%\end{figure}

%For data rate 200Gbps over typical 50cm chip to chip trace, the on-line 8-PAM shaping scheme achieves overall shaping gain of 8.55 dB compared to 4-PAM un-shaped system. This gain is equivalent to a gain of 1.42 bit in the ENOB of the ADC.
%For this working point, the achieved gain is only 3dB less than the gain predicted by the UB as will be explained in Section \ref{Results}. This gap in the gain will be analyzed and explained in Section \ref{Results}.

%===============================================================================
\section{System Model}
\label{sm}
\begin{figure}[ht]
\centering
\includegraphics[width=3in]{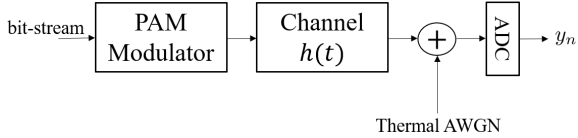} 
\caption{System model.}
\label{block_diagram}
\end{figure}

%===============================================================================
A typical communication link may be adequately described by the model shown in Fig. \ref{block_diagram}. 
%For simplicity we approximate the channel as a discrete channel at the symbol rate. This approximation is justified since the channels at hand have high attenuation at high frequencies which decreases the aliasing effects.
Let $\boldsymbol{X}=(-Q+1,-Q+3,...,Q-3,Q-1)$ be one dimensional $Q$-PAM constellation with cardinality $|\boldsymbol{X}|=Q$,
% i.e, $\boldsymbol{X}=\sqrt{Q}\cdot \boldsymbol{X}/\left\|\boldsymbol{X}\right\|_2$, and $\left\|\cdot\right\|_2$ denotes the $l^2$ norm.
and let ${x}^{N}\triangleq (x_0,...,x_{N-1})$ be a frame of $N$ symbols where $x_n \in \boldsymbol{X}$ $\forall n$.
%From here on, we will use the notation ${x^j}\triangleq (x_{0},x_{1},...,x_{j-1})$.
The frame ${x}^{N}$ is transmitted in symbol rate $f_s$ symbols/sec through a noisy channel with an impulse response ${h(t)}$ and sampled by ADC every $1/f_s$ seconds. The resulting sampled signal at the ADC output is given by
\begin{equation}
\label{discrete_waveform}
y_n =\sum_{i=0}^{L-1}h_i x_{n-i} + z_n+\eta_n=r_n+z_n+\eta_n
\end{equation}
where $z_n$ and $\eta_n$ are two independent sources of one dimensional white Gaussian noise with variance $N_{0}/2$ and $N_{A}/2$, respectively, $L$ is the channel span in symbol periods units, and $r_n=\small\sum_{i=0}^{L-1}h_i x_{n-i}$ is the sampled signal at the $n$-th time step. The noise $z_n$ is receiver thermal noise and $\eta_n$ is additional noise caused by the ADC as a result of the quantization process and distortion, approximated as AWGN. This approximation is justified since quantization noise in practical ADCs rarely have uniform quantization noise. The noise in practical ADCs is influenced by inaccuracies, non-linearities, clock jitter, and thermal noise inside ADC which overall can be approximated as white gaussian noise sources \cite{real_ADC3}.
%The instantaneous power of the received signal is $p_n = |r_n|^2$ and the PAPR at the ADC input is the ratio between the peak power $p_{peak}$ and the average power $P_r=E\{p_n\}$, where $p_{peak}$ is defined as the value of $p_n$ which is exceeded with probability $\epsilon$. In this paper we use $\epsilon=10^{-4}$. In addition, we normalized constellation points so that average uniform transmission power is 1.

The instantaneous power of the received signal is $p_n = |r_n|^2$ and the Signal to Noise Ratio (SNR) is defined by (\ref{SNR}) where $P_r=E\{p_n\}$ is the average power of the signal, and $E\{\cdot\}$ denotes the statistical averaging.

\begin{equation}
\label{SNR}
\small SNR\triangleq \frac{2P_r}{N_{A}+N_0}
\end{equation}
The PAPR at the ADC input is the ratio between the peak power $p_{peak}$ and the average power $P_r$, where $p_{peak}$ is defined as the value of $p_n$ which is exceeded with probability $\epsilon$. In this paper we use $\epsilon=10^{-4}$. In addition, we normalized constellation points so that average uniform transmission power is 1. 

%\begin{equation}
%\label{powerOfsignal}
%p_n = |r_n|^2 =\bigg|\sum_{i=0}^{L-1}h_i x_{n-i}\bigg|^2 
%\end{equation}

% Under the condition $N_{0}<<\gamma$,
%and the PAPR at the ADC input is the ratio between the peak power $p_{peak}$ and the average power $P_r=E\{p_n\}$, where $p_{peak}$ is defined as the value of $p_n$ which is exceeded with probability $\epsilon$. In this paper, we use $\epsilon=10^{-4}$.% i.e.,
%\begin{equation}
%\label{PAPRRX}
%PAPR = \frac{p_{peak}}{P_r} 
%\end{equation}

The Signal to Noise and Distortion Ratio (SNDR) and the ENOB of an ADC device are defined by \cite{IEEEADC}
\begin{equation}
\label{SNR_ADC}
\small SNDR\triangleq \frac{2P_r}{N_{A}}
\end{equation}

\begin{equation}
\label{ENOB}
\small ENOB \triangleq \frac{10\cdot log_{10}(SNDR\cdot PAPR)-4.76}{6}
\end{equation}

Since shaping for PAPR reduction allows an equivalent increment in the average power of the received signal, the overall shaping gain $G_T$ is the sum of the PAPR and SNDR gains (if denoted in dB), given constant ratio between the transmitted average power $P_t=E\{|x_n|^2\}$ and thermal noise $N_0$. This ratio is denoted by Transmitter Signal to Thermal Noise Ratio (TSTNR).

A typical wireline channel, with a causal continuous time impulse response, could be approximated as \cite{channel} 

\begin{equation}
\label{channelimpulse}
h(t)=\frac{A}{\sqrt{(t_0+t)^3}e^{\frac{\pi A^2}{(t_0+t)}}}\circledast \frac{2B}{\pi ((t_0+t)^2+B^2)},   t\geq 0
\end{equation}
where A and B are positive constants that determine the relaxation time of the response and $t_0 \geq 0$ is a parameter that determines the first sample of $h(t)$. In this paper, we use $t_0=7.7\cdot10^{-13} sec$, $A=10^{-6} \sqrt{sec}$ and $B=8.8\cdot10^{-12} sec$, which are typical values of a microstrip trace of length 50 cm used for communicating between two chips \cite{channel}. The discrete sampled impulse response is therefore $h_n=h(t_0+n/f_s)$. 
The symbol rates we use in this paper are $f_s = 112$ Gsymbol/sec and $f_s = 224$ Gsymbol/sec. The resulting voltage gain normalized sampled impulse responses are

\begin{equation}
\label{cha}
\begin{split}
h_A & = \{0.13, 0.19, 0.14, 0.09, 0.07, 0.05, 0.037, 0.031, 0.025,\\ 
&  0.02, 0.016, 0.014, 0.013, 0.012, 0.011, 0.01, 0.009, 0.008,\\
&  0.0075, 0.0072, 0.0065, 0.0071, 0.0057, 0.0055, 0.0044,\\	
&  0.0044, 0.0033, 0.0033, 0.0032, 0.0029\}.
\end{split}
\end{equation}

%0.069, 0.1, 0.11, 0.098, 0.08, 0.06, 0.05, 0.04, 0.038, 0.032, 0.028, 0.024, 0.021, 0.019, 0.017, 0.015, 0.014, 0.013, 0.0118, 0.0108, 0.01, 0.0092, 0.0086, 0.008, 0.0075, 0.007, 0.0066, 0.0062, 0.0058, 0.0055, 0.0052, 0.00498, 0.00474, 0.00451, 0.00429, 0.0041, 0.0039, 0.0037, 0.0036, 0.0034

\begin{equation}
\label{chb}
\begin{split}
h_B & = \{0.069, 0.1, 0.11, 0.098, 0.08, 0.06, 0.05, 0.04, 0.038,\\ 
&  0.032, 0.028, 0.024, 0.021, 0.019, 0.017, 0.015, 0.014,\\
&  0.013, 0.0118, 0.0108, 0.01, 0.0092, 0.0086, 0.008, 0.0075, \\	
& 0.007,  0.0066, 0.0062, 0.0058, 0.0055, 0.0052, 0.00498,\\
&  0.00474, 0.00451, 0.00429, 0.0041, 0.0039, 0.0037,\\
& 0.0036, 0.0034, 0.0037, 0.0034, 0.0029, 0.0028,\\
& 0.0028, 0.0025, 0.0023, 0.0023, 0.002, 0.0017\}.
\end{split}
\end{equation}
The impulse responses (\ref{cha}) and (\ref{chb}) are denoted Channel-A and Channel-B, respectively. Note that Channel-A and Channel-B span over $L = 30$ and $L=50$ symbols, respectively.

\section{Implementation}
\label{Implementation}
%===============================================================================

%We now describe a practical shaping scheme to control the PAPR of the signal at the channel output. As discussed in Section \ref{intro}, the shaping scheme uses precoding at transmitter and M-BCJR module at receiver, combined with an outer Error Correcting Code (ECC).% A general block diagram of the system is illustrated in Fig. \ref{blockD}.

%\begin{figure}[ht]
%\centering
%\includegraphics[width=3.4in]{TXRX.png} 
%\caption{(a) Transmitter block diagram. (b) Receiver block diagram.}
%\label{blockD}
%\end{figure}

\begin{figure}[ht]
\centering
\includegraphics[width=3in]{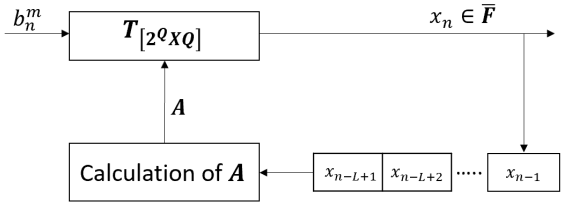} 
\caption{The precoding process.}
\label{precoder}
\end{figure}

A binary information stream $\boldsymbol{u}$ is firstly encoded by an Error Correcting Code (ECC) into a code word in rate $R$ bit/symbol.
In every time step $n$, the precoder maps $m=\log_2(Q)$ coded bits ${b}^{m}_n\triangleq (b_{n0}, b_{n1},..b_{n(m-1)})$ to a symbol $x_n$ that satisfies a peak power constraint $p_n\leq\gamma$.
\begin{comment}
The precoder firstly calculates an indicator vector $\boldsymbol{A}=({A_0},{A_1},...,{A_{Q-1}})$ of the constellation $\boldsymbol{X}$ according to  
\begin{equation}
A_i=\mathbbm{1}_{\boldsymbol{F}}({x_i})=
    \begin{cases}
      0, & x_i\in \boldsymbol{F} \\
      1, & x_i\in \boldsymbol{\overline{F}}
    \end{cases}
    , i=0,1,...,Q-1
    \label{indicators_set}
\end{equation}
where 
\begin{equation}
\label{Fcalc}
   \boldsymbol{\overline{F}}=\bigg \{x \in \boldsymbol{X} \colon \bigg|h_0x+\sum_{i=1}^{L-1}h_i x_{n-i}\bigg|^2 \leq \gamma \bigg \} 
 % \\
\end{equation}
and $\boldsymbol{{F}}= \{x \in \boldsymbol{X} \colon x \notin \boldsymbol{\overline{F}}\}$. Note that $\boldsymbol{F} \cap \boldsymbol{\overline{F}}=\emptyset$ and $\boldsymbol{F} \cup \boldsymbol{\overline{F}}=\boldsymbol{X}$. 
The calculation of $\boldsymbol{A}$ (\ref{indicators_set}) is repeated in every step $n$ by substituting the last $L-1$ transmitted symbols $(x_{n-1},x_{n-2},...,x_{n-L+1})$ to (\ref{Fcalc}). 
Then, according to a mapping table $\boldsymbol{T}$ and the set $\boldsymbol{A}$ in time step $n$, the bits ${b}^{m}_n$ are uniquely map to a symbol $x_n \in \boldsymbol{\overline{F}}$. 
Note, the size of $\boldsymbol{T}$ is $2^Q$-by-$Q$ (it does not depend on the channel length $L$).
The precoder operation is summarized by the block diagram illustrated in Fig. \ref{precoder}.
\end{comment}
%%%%%%%%%%%%%%%%%%%%%%%%%%%%%%%%%%%%%%%%%%%%%
To do so, the precoder firstly calculates the forbidden symbols for transmission at step $n$ (symbols that would yield $p_n>\gamma$) according to the \textit{channel state} ${s}_n$, where ${s}_n$ is defined as the last $L-1$ transmitted symbols $(x_{n-1},x_{n-2},...,x_{n-L+1})$.
The sets of the forbidden and non-forbidden symbols are denoted by $\boldsymbol{F}$ and $\boldsymbol{\overline{F}}$, respectively, where $\boldsymbol{F} \cap \boldsymbol{\overline{F}}=\emptyset$ and $\boldsymbol{F} \cup \boldsymbol{\overline{F}}=\boldsymbol{X}$. 
The calculation of $\boldsymbol{F}$ is preformed according to (\ref{Fcalc}), and $\boldsymbol{\overline{F}}= \{x \in \boldsymbol{X} \colon x \notin \boldsymbol{F}\} $.

\begin{equation}
\label{Fcalc}
   \boldsymbol{F}=\bigg \{x \in \boldsymbol{X} \colon \bigg|h_0x+\sum_{i=1}^{L-1}h_i x_{n-i}\bigg|^2 > \gamma \bigg \} 
 % \\
\end{equation}
Let us define the indicator vector $\boldsymbol{A}=({A_0},{A_1},...,{A_{Q-1}})$ of the constellation $\boldsymbol{X}$ as
\begin{equation}
A_i=\mathbbm{1}_{\boldsymbol{F}}({x_i})=
    \begin{cases}
      0, & x_i\in \boldsymbol{F} \\
      1, & x_i\in \boldsymbol{\overline{F}}
    \end{cases}
    , i=0,1,...,Q-1
    \label{indicators_set}
\end{equation}
%The calculation of $\boldsymbol{A}$ (\ref{indicators_set}) is repeated in every step $n$ by substituting the channel state ${s_n}$ to (\ref{Fcalc}). 
%Then, 
According to a mapping table $\boldsymbol{T}$ and the set $\boldsymbol{A}$ in time step $n$, the bits ${b}^{m}_n$ are uniquely map to a symbol $x_n \in \boldsymbol{\overline{F}}$. 
Note, the size of $\boldsymbol{T}$ is $2^Q$-by-$Q$ (it does not depend on the channel length $L$).
The precoder operation is summarized by the block diagram illustrated in Fig. \ref{precoder}.
%%%%%%%%%%%%%%%%%%%%%%%%%%%%%%%%%%%%%%%%%%%%%%%%%%%%%%%%%%%%%%%%%%%%%%%%%

The mapping table $\boldsymbol{T}$ is constructed according to the following.
If $\boldsymbol{F}=\emptyset$, $\boldsymbol{\overline{F}}=\boldsymbol{X}$ and the symbols bit labeling is the Gray labeling. Otherwise, all the bit labels of the symbols in $\boldsymbol{F}$ cannot be used, and should be assigned to a corresponding symbols from $\boldsymbol{\overline{F}}$. Each label from $\boldsymbol{F}$ should be assigned to a symbol from $\boldsymbol{\overline{F}}$ such that the hamming distance between the labels is minimal. In case of several choices of symbols from $\boldsymbol{\overline{F}}$, with the same minimal hamming distance, the symbol with the lowest Euclidean distance is chosen. The different bits among the common labels of a symbol are equivalent to erasure (could be zero or one since they are unknown to the receiver). The mapping table $\boldsymbol{T}$ of a 4-PAM constellation, $\boldsymbol{X}=(-3,-1,1,3)$, is presented in Table \ref{all_mapping_tables}. The row index is the decimal representation of the binary set $\boldsymbol{A}$ (i.e., if for example $\boldsymbol{A}=(0,1,1,1)$ and ${b}^{m}_n=(1,1)$, the transmitted symbol is $x_n=\boldsymbol{T}_{73}=3$). 
\begin{table} [H]
\begin{center}
    \begin{tabular}{| c | c | c | c | c |}
    \hline
       &10 & 00 & 01 & 11  \\ \hline
    0& - & - & - & - \\ \hline 
    1& 3 & 3 & 3 & 3 \\ \hline 
    2& 1 & 1 & 1 & 1 \\ \hline 
    3& 3  & 1 & 1 & 3 \\ \hline 
    4& -1 & -1 &-1  &-1  \\ \hline 
    5& -1 & -1 & 3 & 3 \\ \hline 
    6& -1 & -1 & 1 & 1 \\ \hline 
    7& -1 & -1 & 1 & 3 \\ \hline 
    8& -3& -3 &-3  &-3  \\ \hline 
    9& -3 & -3 & 3 & 3 \\ \hline 
    10& -3 & -3 & 1 & 1 \\ \hline 
    11&-3 & -3 & 1 & 3 \\ \hline 
    12& -3& -1 & -1 & -3 \\ \hline 
    13& -3& -1 & -1 & 3 \\ \hline 
    14&-3 & -1 & 1 & 1 \\ \hline 
    15& -3 & -1 &1  & 3 \\ \hline 
    \end{tabular}
    \caption{Mapping table $\boldsymbol{T}$ of a 4-PAM constellation.}
    \label{all_mapping_tables}
\end{center}
\end{table}

The signal $x^N$ at the precoder output is a Markov process. The $Q^{L-1}$ distinct states of the Markov process are indexed by $i\in \mathbb{Z}$, $i=0,1,...Q^{L-1}-1$. Since $\Pr(s_n=j|{s_{n-1}}=i)=\Pr(j|i)$ $\forall n$ then, the transmission is a stationary time-homogeneous Markov chain, and the transition between channel states is uniquely defined by a symbol $x_{ij}\in \boldsymbol{X}$ i.e., $\Pr(j|i)=\Pr(x_{ij}|i)$ where $x_{ij}$ is the symbol that causes a transition from state $i$ to state $j$.

At the receiver side we used a modified M-BCJR algorithm which computes online the states probabilities of the infinite-state Markov process.
The M-BCJR algorithm \cite{MBCJR} computes ${\zeta_{ij}}_n\triangleq\Pr(s_{n-1}=i;s_{n}=j;{y}^{N})$ for all $0< n \leq N-1$ and for $M$ states with the highest metrics at step $n-1$. 
Next, $m$ Log Likelihood Ratios (LLR), $\Lambda(b_{nl})$, $0\leq l \leq m-1$, are computed for each noisy symbol $y_n$ according to

\begin{figure}[ht]
\centering
\includegraphics[width=2.5in]{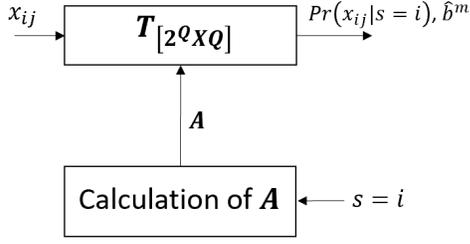} 
\caption{Trellis branch probability online calculation process.}
\label{precoderBCJR}
\end{figure}

\setlength{\arraycolsep}{0.0em}
\footnotesize
\begin{equation}
\begin{split}
{\Lambda(b_{nl})} & =\log\bigg(\sum_{(i,j)}\frac{\sum_{x_{ij}:\hat{b}_l=0}\zeta_{ij_n} + \sum_{x_{ij}:\hat{b}_l=X}\zeta_{ij_n}\cdot \Pr(b_{nl}=0) }{\sum_{x_{ij}:\hat{b}_l=1}\zeta_{ij_n} + \sum_{x_{ij}:\hat{b}_l=X}\zeta_{ij_n}\cdot \Pr(b_{nl}=1) }\bigg)\\
& -{\Lambda^e(b_{nl})}
\end{split}
\label{BCJRllr}
\end{equation}
\normalsize
where the bit label of the symbol $x_{ij}$ is denoted by $\hat{b}^m$ and the ambiguous bits in the bit label are denoted by X.
The bit probabilities $\Pr(b_{nl}=0)$ and $\Pr(b_{nl}=1)$ are calculated from $\Lambda^e(b_{nl})$, which is the extrinsic LLR from the code decoder. %If ECC is not employed then all $\Lambda^e(b_{nl})$ are zero i.e., $\Pr(b_{nl}=0)=\Pr(b_{nl}=1)=0.5$. 
The calculation (\ref{BCJRllr}) requires, for each $x_{ij}$, both $\hat{b}^m$ and the trellis branch probability $\Pr(x_{ij}|{s}=i)$.
In uniform transmission, $\Pr(x_{ij}|{s}=i)=1/Q$ $\forall i,j$ and the symbols bit label is the Gray labeling in all states. 
However, in the suggested shaping scheme, $\Pr(x_{ij}|{s}=i)$ and $\hat{b}^m$ depend on the state $i$, $\gamma$, and ${h}^{L}$. 
%The M-BCJR therefore calculates the set $\boldsymbol{A}$ (\ref{indicators_set}), for each survivor state $i$ at time step $n$. Once $\boldsymbol{A}$ is known, $\hat{b}^m$ and $\Pr(x_{ij}|{s}=i)$ are calculated according to the same mapping table $\boldsymbol{T}$ used at the transmitter.
Calculation of these metrics is preformed according to the process illustrated in Fig. \ref{precoderBCJR}.% The mapping table $\boldsymbol{T}$ is the exact same table used at transmitter.

%We compare the additional complexity required by the online shaping scheme to uniform transmission with a TE in the receiver. Such a system does not use a precoder in transmitter but, uses M-BCJR module in receiver. According to (\ref{Fcalc}), the precoder adds $L$ multiplication and $L-1$ additions operations per time step in the transmitter side. 
%In the receiver side, the calculation of the sets $\boldsymbol{F}$ and $\boldsymbol{\overline{F}}$, according to (\ref{Fcalc}), does not add complexity. Hence, the  symbol bit label and the branch metrics  online calculation does not add complexity to the M-BCJR.

The LLR values ${\Lambda}^{mN}$ at the BCJR output can be used as an a priory input to a ECC decoder. In each iteration, the decoder produces extrinsic LLR values $({\Lambda^e})^{mN}$ which are used as an a priory input to the BCJR module, which in turn calculates new extrinsic LLRs which are sent back to the code decoder. After a pre-determined number of iterations has reached, the bit estimations $\boldsymbol{\hat{u}}$ are determined by performing hard decision on the decoder LLR values $({\Lambda^e})^{mN}$.
Initially, all $({\Lambda^e})^{mN}$ are set to 0.

%\subsection{Combining With a Code}
 
%At the transmitter, a binary information stream $\boldsymbol{u}$ is firstly encoded, with a rate $R$ bit/symbol, by an ECC code into a code word. Any iteratively decodable code that handles erasures can be used. We used Turbo code \cite{Turbo} which possesses excellent capability of handling erasures.
%The code word is then divided into $m$-tuples ${{b}^{m}}$. Then, at time instance $n$, the precoder maps each ${{b}^{m}}_n$ into a symbol $x_n$ using the scheme illustrated in Fig. \ref{precoder}.
%At the receiver side, the LLR values ${\Lambda}^{mN}$ at the BCJR output can be used as an a priory input to a ECC decoder. In each iteration, the decoder produces extrinsic LLR values $({\Lambda^e})^{mN}$ which are used as an a priory input to the BCJR module, which in turn calculates new extrinsic LLRs which are sent back to the code decoder. After a pre-determined number of iterations has reached, the bit estimations $\boldsymbol{\hat{u}}$ are determined by performing hard decision on the decoder LLR values $({\Lambda^e})^{mN}$.
%Initially, all $({\Lambda^e})^{mN}$ are set to 0.

%As suggested in \cite{ReducedBCJR}, the decoding complexity can be reduced by saturating the BCJR extrinsic LLRs (\ref{BCJRllr}) according to $-\mu \leq {\Lambda(b_{nl})}\leq \mu$ where $\mu$ is parameter which is optimized to yield the best Bit Error Rate (BER) performance. 

%===============================================================================
 \section{Theoretical Analysis}
 \label{UB}
 %===============================================================================
%In this section we derive an UB on the RPQNR, given rate and TSTNR. Since the RPQNR is the sum of RSQNR and PAPR (if denoted in dB), an UB on each of these metrics is separately derived. The UB on the RPQNR is the sum of both UBs. 
%The motivation to this section stems from the following two questions \romannum{1}) What is the lowest required SNDR given rate and TSTNR? and \romannum{2}) What is the theoretical shaping gain in a given $\gamma$, rate, and TSTNR? Section \ref{rsqnr_ub} answers the first question by introducing a Lower-Bound (LB) on SNDR given rate and TSTNR. It is well known that in this case the optimal one-dimensional symbols distribution is Gaussian. We optimized the Power Spectral Desnsity (PSD) of the Gaussian distribution such that the achievable rate is maximized. This LB can be used for upper-bounding the SNDR gain by comparing between the LB and SNDR of a flat PSD (i.i.d distribution) at a given rate. 
%In Section \ref{papr_ub} we estimate the receiver PAPR of a peak constrained transmission. We then use this estimation, together with the UB on the SNDR gain, to derive the theoretical shaping gain.

This section aims to study the achievable theoretical gains using shaping. In Section \ref{rsqnr_ub} we derive a Lower-Bound (LB) on the SNDR given rate and TSTNR. It is well known that in this case the optimal one-dimensional symbols distribution is Gaussian. We optimized the Power Spectral Desnsity (PSD) of the Gaussian distribution such that the achievable rate is maximized. This LB can be used for upper-bounding the SNDR gain, by comparing between the LB and the SNDR of a flat PSD (i.i.d distribution) at a given rate.
In Section \ref{papr_ub} we estimate the receiver PAPR of a peak constrained transmission. We then use this estimation, together with the UB on the SNDR gain, to derive the theoretical shaping gain.

\subsection{Upper Bound for Infinite Constellation}
\label{rsqnr_ub}

The channel capacity can be tightly approximated by $C=lim_{N\to \infty}C_N$ \cite{isicapacity} where

\begin{equation}
\label{capacity}
C_N \triangleq \frac{1}{2N}\sum_{i=0}^{N-1} \log\bigg(1+\frac{2q_i|H_i|^2}{N_A+N_0}\bigg),
\end{equation}
${q}^{N}$ are the energy spectral components of a Gaussian input process $g^{N}$, and ${H}^{N}$ is $N$-points Discrete Fourier Transform (DFT) of the channel impulse response ${h}^{L}$. %The $i$-th energy spectral component of the received process is $q_i|H_i|^2$.
UB on the achievable rate given SNDR and TSTNR is found by optimizing the channel capacity (\ref{capacity}) under the following constraints
 
 \begin{equation}
\label{pr_const}
\begin{aligned}
& \frac{1}{N}\sum_{i=0}^{N-1}q_i|H_i|^2 \leq K \\
& \frac{1}{N}\sum_{i=0}^{N-1}q_i \leq P 
\end{aligned}
\end{equation}
where $K, P \in \mathbb{R}_{\geq 0}$ are the constraints on receiver and transmitter average power, respectively.
%UB on SNDR is found by optimizing the channel capacity under the following constraints
%\begin{equation}
%\label{pr_const}
%\begin{aligned}
%& \frac{1}{N}\sum_{i=0}^{N-1}q_i|H_i|^2 \leq K \\
%& \frac{1}{N}\sum_{i=0}^{N-1}q_i \leq P 
%\end{aligned}
%\end{equation}
%where ${q}^{N}$ are the energy spectral components of the input process ${x}^{N}$, $K, P \in \mathbb{R}_{\geq 0}$ are the constraints on receiver and transmitter average power, respectively, and ${H}^{N}$ is $N$-points Discrete Fourier Transform (DFT) of the channel impulse response ${h}^{L}$. %The $i$-th energy spectral component of the received process is $q_i|H_i|^2$.
%The channel capacity is given by $C=lim_{N\to \infty}C_N$ \cite{isicapacity} where
%\begin{equation}
%\label{capacity}
%C_N \triangleq \frac{1}{2N}\sum_{i=0}^{N-1} \log\bigg(1+\frac{2q_i|H_i|^2}{N_A+N_0}\bigg)
%\end{equation}
%The constraints (\ref{pr_const}) converges to receiver and transmitter average power for $N\to \infty$. 
%Therefore, the constraints (\ref{pr_const}) and the capacity (\ref{capacity}) are approximations which become more accurate as $N$ increases. 
%Hence, $N$ should be arbitrarily large for a tight approximation. In this paper we used $N=10^6$.

We determine the maximum of $C_N$ subject to the constraints (\ref{pr_const}) by introducing Lagrange multipliers $(\alpha,\beta)$, $\alpha,\beta \geq 0$, and find the maximum of

\begin{equation}
\label{LAGRA}
\begin{split}
J&=\sum_{i=0}^{N-1}\log\bigg(1+\frac{2q_i|H_i|^2}{N_A+N_0}\bigg)-\alpha\bigg(\sum_{i=0}^{N-1}q_i|H_i|^2-K\bigg)\\
&-\beta\bigg(\sum_{i=0}^{N-1}q_i-P\bigg)
\end{split}
\end{equation}
We get
\begin{equation}
\label{LAGRAd}
\frac{\partial J}{\partial q_i} = \frac{2|H_i|^2}{2q_i|H_i|^2+N_A+N_0}-\alpha|H_i|^2-\beta=0
\end{equation}
Solving (\ref{LAGRAd}) for $q_i$ yields  
\begin{equation}
\label{opt_s}
q^{o}_i = \max \Bigg[0, \frac{1}{\alpha|H_i|^2+\beta}-\frac{N_A+N_0}{2|H_i|^2} \Bigg]
\end{equation}
Capacity is achievable if the input sequence ${g}^{N}$ is a Gaussian process with energy spectral components $({q^{o}})^{N}$ and the multipliers $\alpha$ and $\beta$ are chosen such that the constraints (\ref{pr_const}) are satisfied.
% i.e.,
%\setlength{\arraycolsep}{0.0em}
%\footnotesize
%\begin{equation}
%\label{lagr_calc}
%\begin{aligned}
%& \frac{1}{N}\sum_{i=0}^{N-1}q_i^o|H_i|^2 =\frac{1}{N}\sum_{i=0}^{N-1}\max %\Bigg[0, \bigg(\frac{|H_i|^2}{\alpha|H_i|^2+\beta} %-\frac{N_A+N_0}{2}\bigg)\Bigg]= K \\
%& \frac{1}{N}\sum_{i=0}^{N-1}q_i^o =\frac{1}{N}\sum_{i=0}^{N-1}\max\Bigg[0, %\bigg(\frac{1}{\alpha|H_i|^2+\beta}-\frac{N_A+N_0}{2|H_i|^2}\bigg)\Bigg] = P 
%\end{aligned}
%\end{equation}
%\normalsize
Since we are interested in $({q^{o}})^{N}$ that yields the highest capacity for a given SNDR and TSTNR values, we optimized the capacity (\ref{capacity})  with respect to $K$, while TSTNR and SNDR are kept constants i.e.,

\begin{equation}
\label{optK}
C_o=\max_{K}\bigg(C_N\vert_{\small(\small SNDR,\small TSTNR\small)}\bigg)
\end{equation}
The value of $K$ that maximized (\ref{optK}) is denoted as $K_{o}$.

\begin{figure}[ht]
\centering
\subfloat[]{{\includegraphics[width=1.65in]{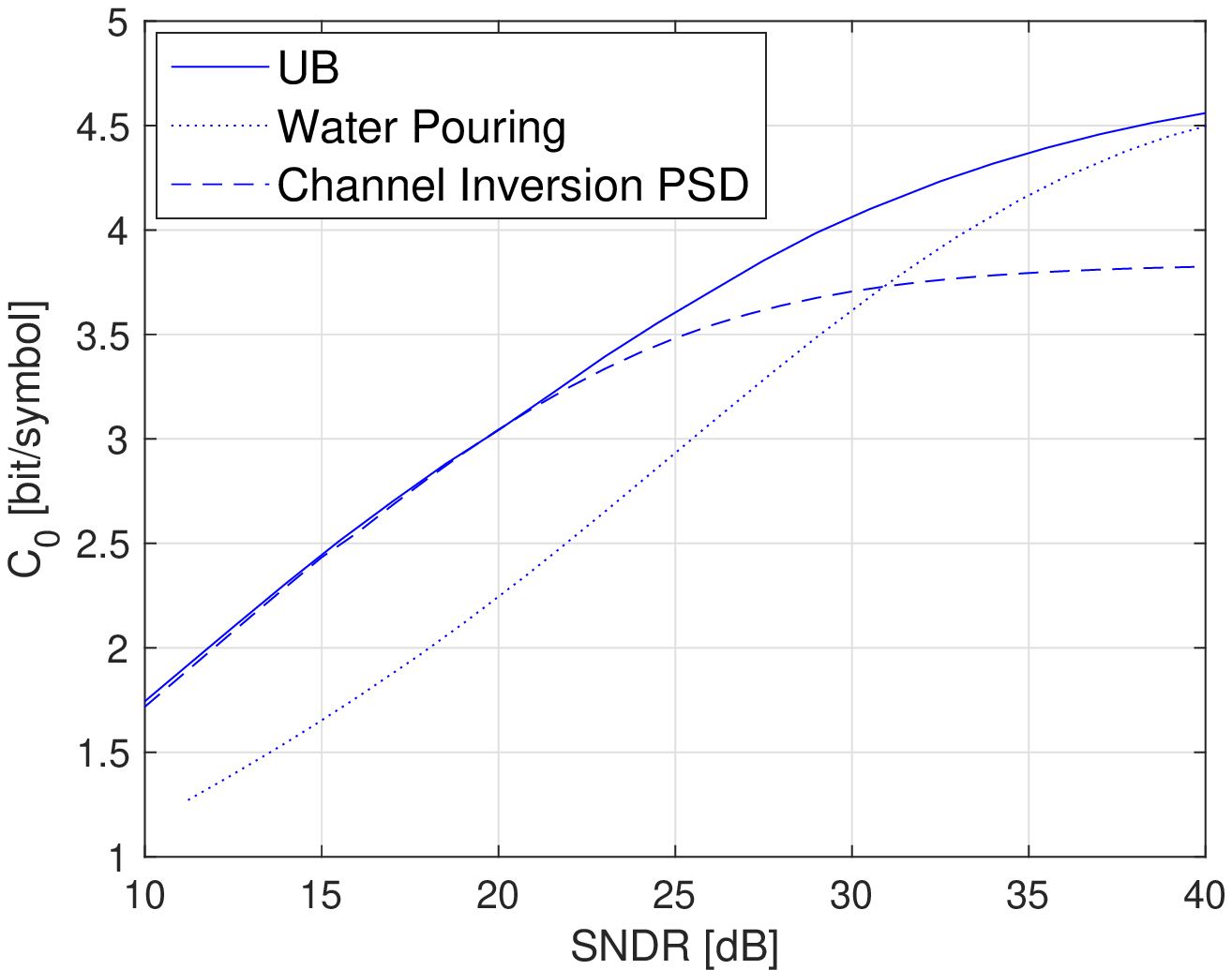} }}
%\qquad
\subfloat[]{{\includegraphics[width=1.65in]{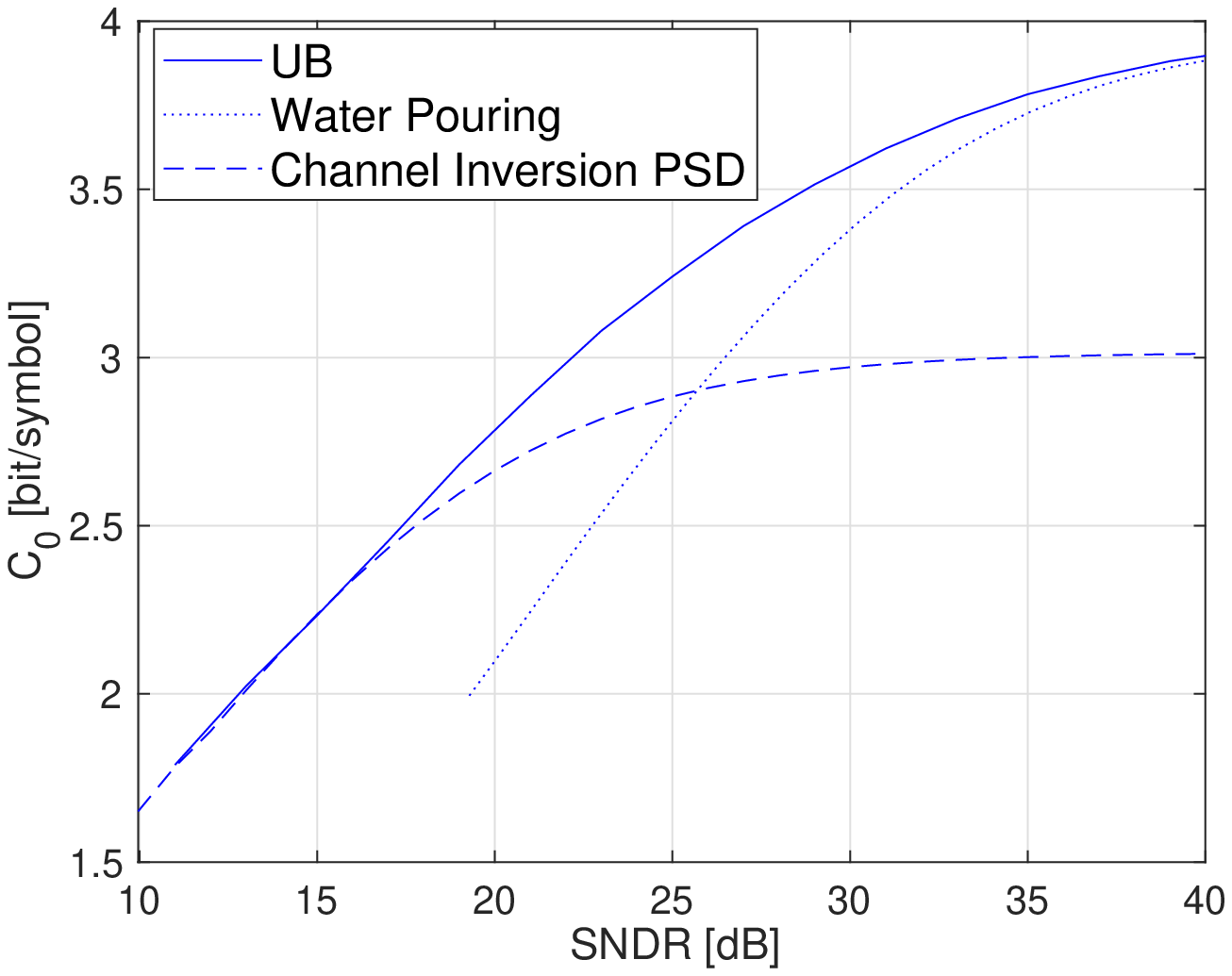} }}
\caption{UB on the achievable rate for Channel-A. (a) TSTNR 45 dB. (b) TSTNR 40 dB.}
\label{UBSTNR45}
\end{figure}

The UB expression (\ref{opt_s}) can be divided to three regions, (a)  $N_A>>N_0$, (b) $N_A<<N_0$, and (c) $N_A \approx N_0$. In the region $N_A>>N_0$, $N_0$ has a negligible influence on the total noise power. The optimization process (\ref{optK}) therefore yields low $K_{o}$ value. The reason is that in low $K$ value, the constraint on the transmitted average power $P$ is not effective since it is already satisfied. Hence, the optimal solution is to invert the channel, which is obtained from (\ref{opt_s}) by setting $\beta=0$. The Lagrange multiplier $\alpha$ is chosen such that the average power constraint at the receiver is kept. 
In the region $N_A<<N_0$, $N_A$ has a negligible influence on the total noise. Since $N_0$ is constant, the optimization process (\ref{optK}) yields high $K$ value. However, the receiver power constraint is not effective in case where $K$ is higher than the average power that would have been obtained at the receiver without any constraint (as it is already met). The optimal solution is therefore reduced to the well-known water-pouring solution, which is obtained from (\ref{opt_s}) by setting $\alpha=0$. The Lagrange multiplier $\beta$ is chosen such that the average power constraint at the transmitter is kept.
In the region $N_A \approx N_0$, both noises influences on the total noise power thus, the optimal solution is given by (\ref{opt_s}). As an example, the channel capacity (\ref{optK}), under the constraints (\ref{pr_const}), was calculated over Channel-A for TSTNR 45 dB and 40 dB.
UB on the rate (or, equivalently, LB on the SNDR) in these TSTNR values is illustrated in Fig. \ref{UBSTNR45}(a) and Fig. \ref{UBSTNR45}(b), respectively.

\subsection{Estimation of Receiver PAPR}
\label{papr_ub}

In an un-shaped transmission, the one-dimensional distribution of each sample at channel output, according to CLT, approaches the Gaussian distribution. In a peak constrained transmission, this distribution could be therefore approximated by the Truncated Gauss (TG) distribution in the region $[-\sqrt{\gamma},\sqrt{\gamma}]$. The probability density function of such distribution is

\begin{equation}
\label{truncated_gauss_pdf}
f(r)=\frac{\exp\Big({-{r^2}/{2\sigma^2}}\Big)}{\sqrt{2\pi\sigma^2}(1+erf(\small\sqrt{{\gamma} / {2\sigma^2}}))}
\end{equation}

\begin{table}[ht]
\centering
    \begin{tabular}{| c |c |c |c|c|}
    \hline
    &\multicolumn{2}{|c|}{Channel-A}&\multicolumn{2}{|c|}{Channel-B} \\
    \hline
     $\gamma$  & TG Gauss & Simulation & TG Gauss & Simulation  
    \\ \hline  
   -16 dB  &4.89 dB &4.9 dB &4.95 dB & 5.15 dB     \\ \hline
    -14 dB &4.96 dB&5.27 dB  & 5.06 dB & 5.3 dB    \\ \hline
   -12 dB & 5.07 dB& 5.3 dB & 5.24 dB & 5.42 dB   \\ \hline
   -10 dB &5.24 dB &5.56 dB  &5.51 dB & 5.63 dB   \\ \hline
   -8 dB & 5.52 dB & 5.75 dB  & 5.95 dB & 6.1 dB   \\ \hline
   -6 dB & 6 dB & 6.07 dB  & 6.65 dB & 6.85 dB   \\ \hline
   -4 dB &6.65 dB &6.63 dB   &7.73 dB & 8.09 dB \\ \hline
   -2 dB &7.7 dB &7.58 dB  & 9.15 dB &  9.6 dB   \\ \hline
    0 dB &10.08 dB &10.03 dB  & 11.17 dB& 11.4 dB   \\ \hline
    \end{tabular}
    \caption{PAPR of TG Gauss and online shaping scheme.}
    \label{papr_comp}
\end{table}

and its PAPR is
\begin{equation}
\label{truncated_gauss_papr}
PAPR_{TG}=\frac{\gamma}{K_{TG}}
\end{equation}

where

\begin{equation}
\label{K_TG}
K_{TG}(\gamma)=\sigma^2-{erf\Big(\small\sqrt{{\gamma}/{2\sigma^2}}\Big)}^{-1}\sqrt{\frac{{2\gamma\sigma^2}}{\pi}}\exp({-\gamma/2\sigma^2}) 
\end{equation}
is the average power and $\sigma^2$ is the un-shaped (i.i.d) received signal average power i.e., $\sigma^2=\sum_{i=0}^{L-1}{h_i}^2$.
%\begin{equation}
%\label{iid_rx_power}
%\sigma^2=\sum_{i=0}^{L-1}{h_i}^2
%\end{equation}

A comparison between (\ref{truncated_gauss_papr}) and the PAPR which yields the online shaping scheme at Channel-A and Channel-B outputs is summarized by Table \ref{papr_comp}, for different $\gamma$ values. It can be seen that indeed (\ref{truncated_gauss_papr}) approximates well the practical PAPR achieved by the online shaping scheme.

\subsection{Theoretical Shaping Gain}
\label{theoretica_GT}

%An estimation of the theoretical shaping gain, given $\gamma$, rate and TSTNR, is found by    

\begin{figure}[ht]
\centering
\subfloat[]{{\includegraphics[width=1.65in]{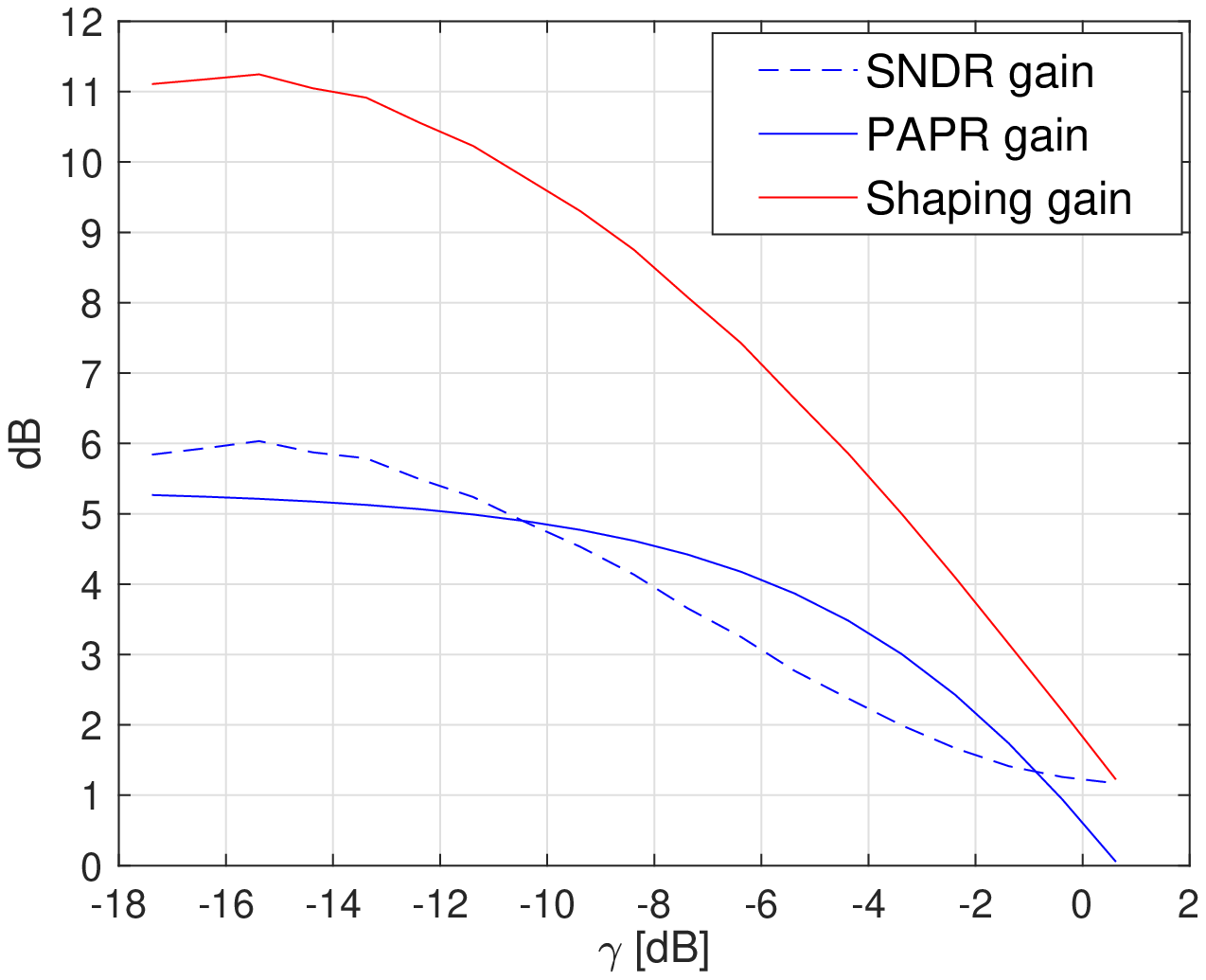} }}
\subfloat[]{{\includegraphics[width=1.65in]{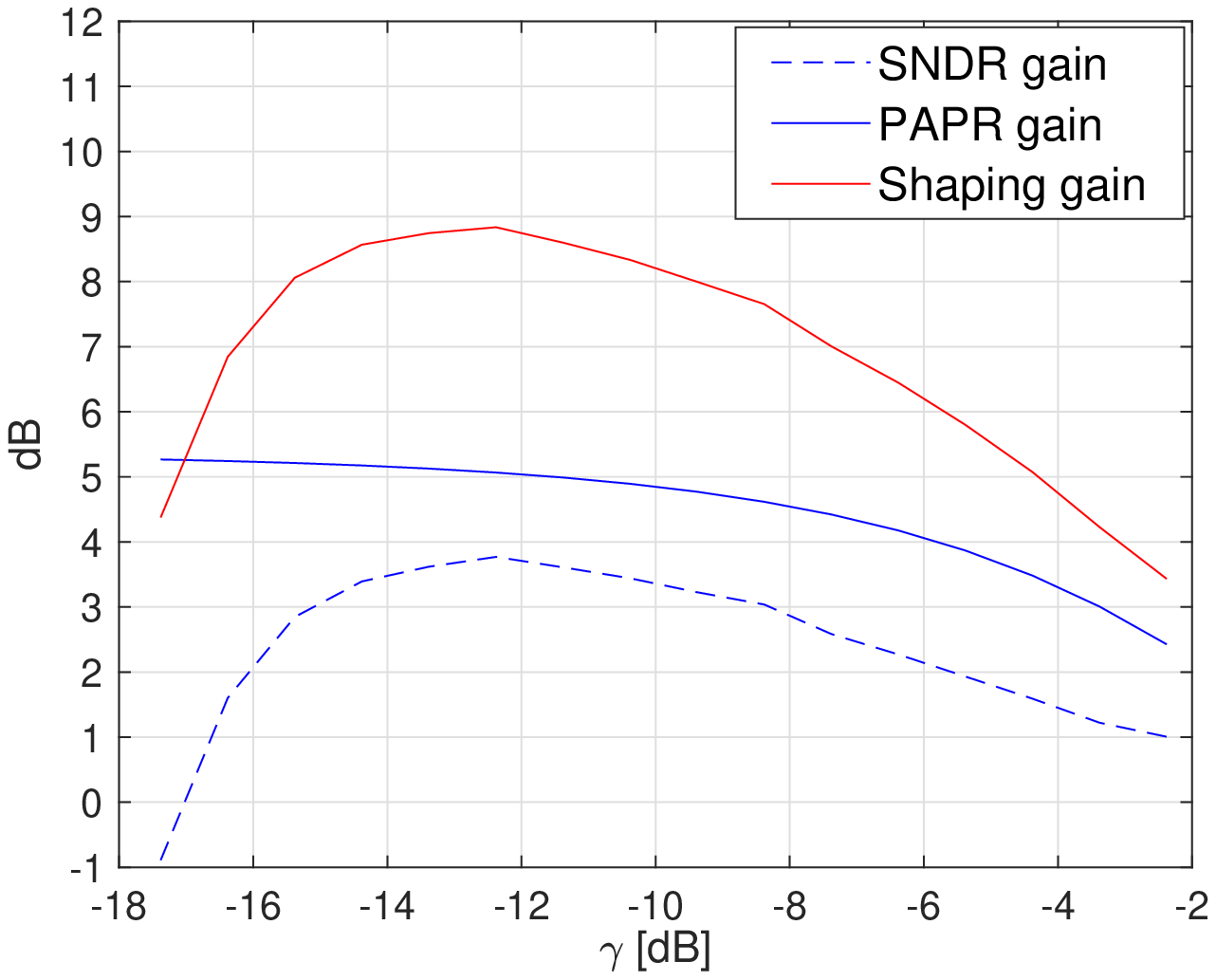} }}
\caption{Relationship between theoretical gains and $\gamma$ in rate 1.8 bits/symbol over Channel-A. (a) TSTNR 40 dB. (b) TSTNR 34 dB.}
\label{THEOR_G_T}
\end{figure}

The PAPR gain in a specified $\gamma$ is found by comparing (\ref{truncated_gauss_papr}) to receiver PAPR of uniform 4-PAM transmission. 
The SNDR gain is found in a specified $\gamma$, rate and TSTNR, by comparing the theoretical SNDR of an un-shaped (i.i.d) Gaussian input distribution and the theoretical SNDR achieved by constraining the receiver power to $K=K_{TG}({\gamma})$. The relationships between the theoretical shaping gains and $\gamma$ in rate 1.8 bits/symbol over Channel-A are demonstrated in Fig. \ref{THEOR_G_T}(a) and Fig. \ref{THEOR_G_T}(b), for TSTNR of 40 dB and 34 dB, respectively.
%An example of such a comparisons in rate 1.8 bits/symbol over Channel-A are demonstrated in Fig. \ref{THEOR_G_T}(a) and Fig. \ref{THEOR_G_T}(b) for TSTNR of 40 dB and 34 dB, respectively. 
It can be seen that the maximal theoretical shaping gains in these cases are 11.65 dB and 8.83 dB, respectively.   
The maximal theoretical shaping gains in rate 1.8 bits/symbol over Channel-A, and the corresponding $\gamma$ values, are summarized in Table \ref{theoretica_GT_vs_tstnr} for several TSTNR values.

\begin{table}[ht]
\centering
    \begin{tabular}{| c |c |c |}
    \hline
      TSTNR & $\gamma$ &  $G_T$ 
    \\ \hline  
   45 dB  & -15 dB &12.34 dB     \\ \hline
    40 dB &-15 dB &11.25 dB \\ \hline
   37 dB & -13 dB& 9.75 dB \\ \hline
   34 dB &-12 dB &8.83 dB    \\ \hline
   31 dB & -10 dB & 7.71 dB     \\ \hline
   29 dB & -7 dB & 6.5 dB   \\ \hline
    \end{tabular}
    \caption{Theoretical shaping gains in rate 1.8 bits/symbol over Channel-A for several TSTNR values.}
    \label{theoretica_GT_vs_tstnr}
\end{table}

%\begin{table}[ht]
%\centering
%    \begin{tabular}{| c |c |c |c|c|}
%    \hline
%    &\multicolumn{2}{|c|}{Channel-A}&\multicolumn{2}{|c|}{Channel-B} \\
%    \hline
%      TSTNR & $\gamma$ &  $G_T$ & $\gamma$& $G_T$
%    \\ \hline  
%   45 dB  & -15 dB &12.34 dB& -20 dB & 14.42 dB       \\ \hline
%    40 dB &-15 dB &11.25 dB &-19 dB & 13.3 dB  \\ \hline
%   37 dB & -13 dB& 9.75 dB &-17 dB & 11.6 dB\\ \hline
%   34 dB &-12 dB &8.83 dB & &   \\ \hline
%   31 dB & -10 dB & 7.71 dB& &    \\ \hline
%   29 dB & -7 dB & 6.5 dB & &  \\ \hline
%    \end{tabular}
%    \caption{Theoretical shaping gains for several TSTNR values.}
%    \label{theoretica_GT_vs_tstnr}
%\end{table}

%===============================================================================
\section{Simulation Results}
\label{Results}
The shaping was applied over 4-PAM and 8-PAM constellations with code rates 0.9 and 0.6, respectively. Hence, the data rate in all systems is $R=1.8$ bits/symbol or, equivalently, 200 Gbps for Channel-A and 400 Gbps for Channel-B.

The code used with all schemes is a standard turbo encoder \cite{Turbo}, made up of two elementary encoders with memory size 4 and the same generator polynomial 37-23 (octal number 37 represents the feed-forward connections and 23 the feedback connections). This code is known to be an optimal code with memory size 4 for various turbo-code rates \cite{optTurbo}.
At the receiver, the number of survivors states we used in the M-BCJR module, in all systems, was $M=16$ states per time step. 
The turbo decoding ran for maximum 12 iterations on block length of 4096 information bits. 
The shaped systems are compared to uniform 4-PAM transmission with TE at the receiver, over the same channel.

\begin{figure}[ht]
\centering
\subfloat[]{{\includegraphics[width=1.65in]{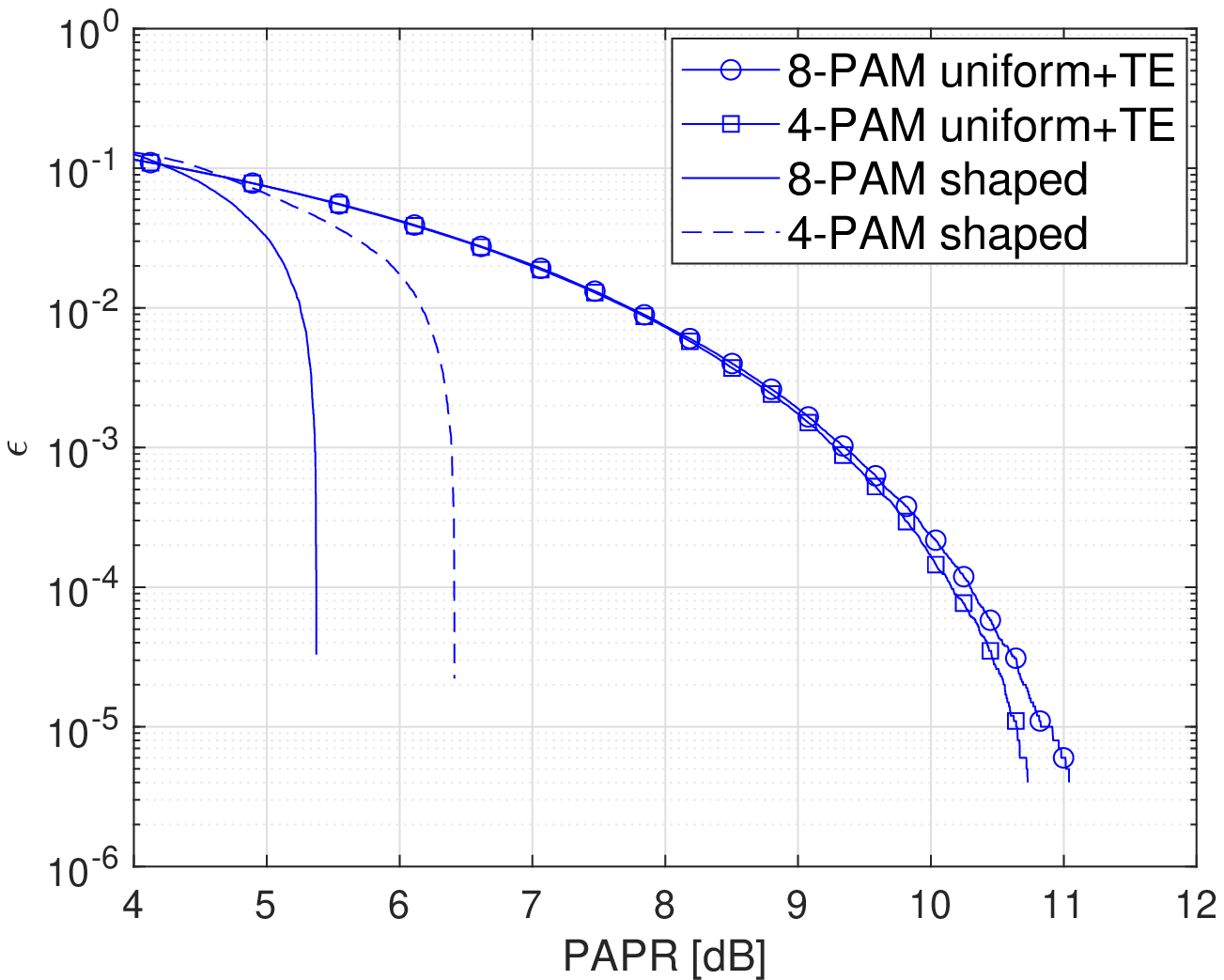} }}
%\qquad
\subfloat[]{{\includegraphics[width=1.65in]{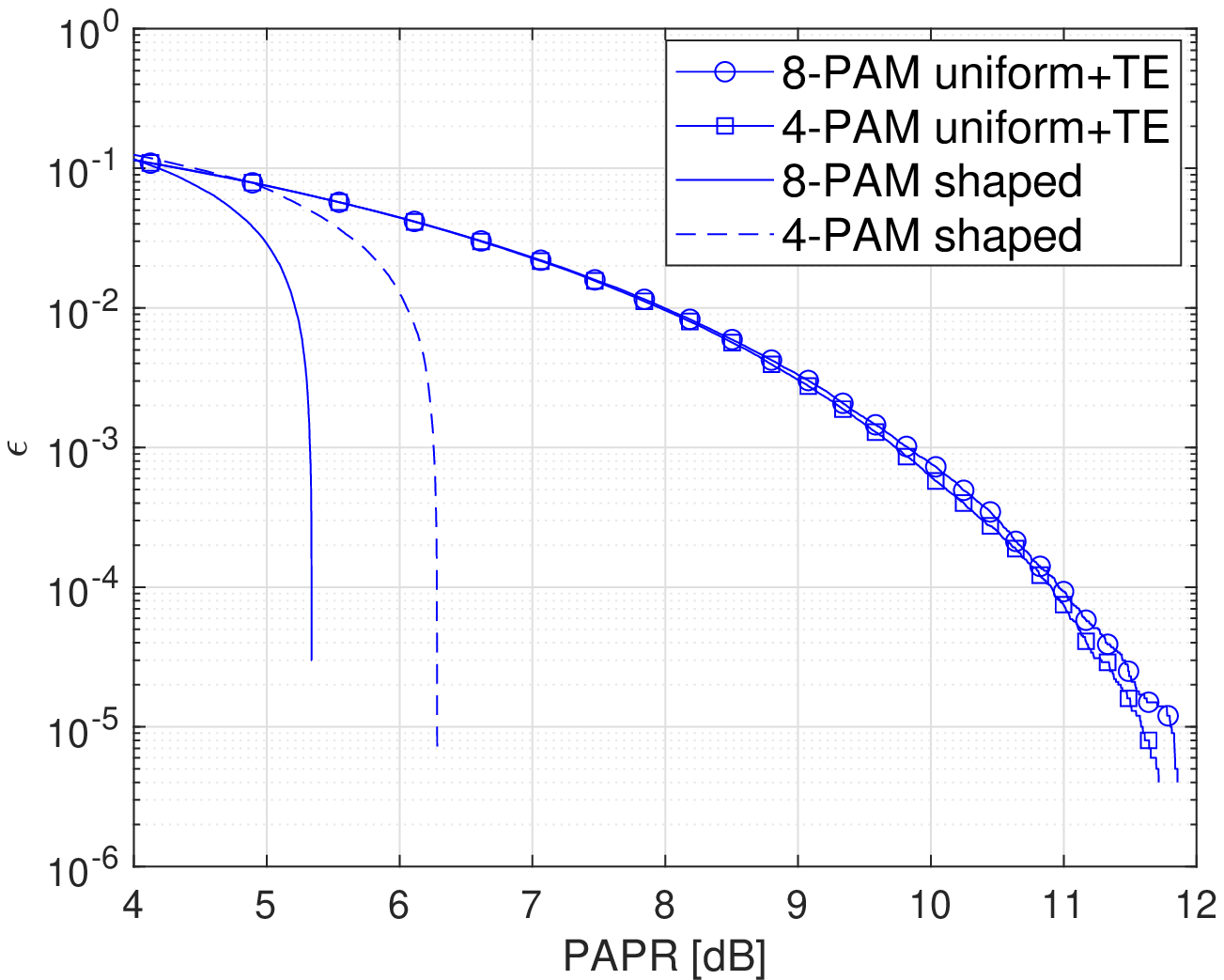} }}
\caption{PAPR distributions at the channel output for TSTNR 40 dB and rate 1.8 bits/symbol. (a) Channel-A. (b) Channel-B.}
\label{PAPRDIST}
\end{figure}

\begin{figure}[ht]
\centering
\subfloat[]{{\includegraphics[width=1.65in]{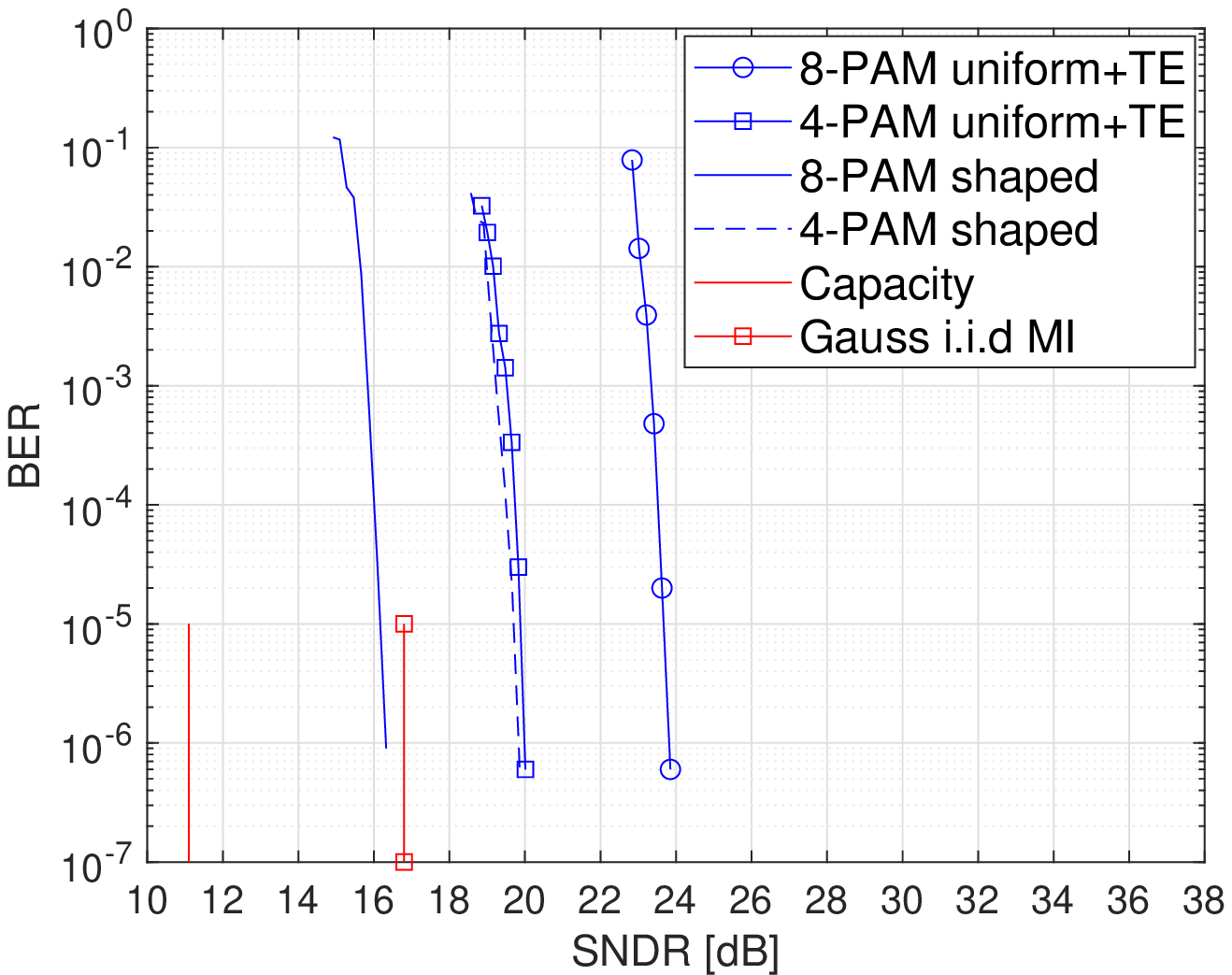} }}
%\qquad
\subfloat[]{{\includegraphics[width=1.65in]{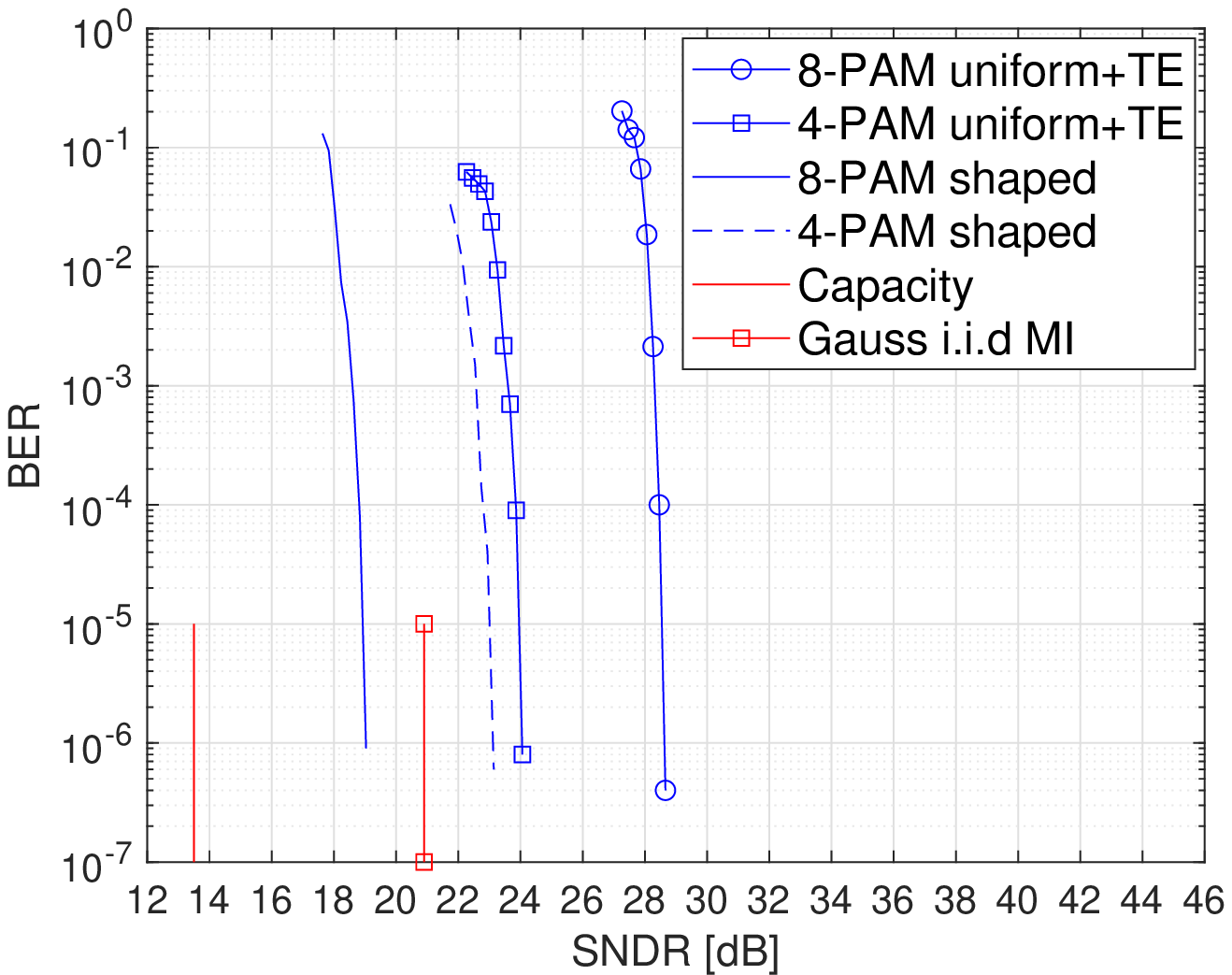} }}
\caption{BER Vs. SNDR for TSTNR 40 dB and rate 1.8 bits/symbol. (a) Channel-A. (b) Channel-B.}
\label{BERPQNRB}
\end{figure}

%\begin{figure}[ht]
%\centering
%\includegraphics[width=2in]{PAPR_RX_V1.eps} 
%\caption{PAPR distributions at Channel-A output.}
%\label{PAPRDIST}
%\end{figure}

%\begin{figure}[ht]
%\centering
%\includegraphics[width=2in]{RSDNRBER_TSTNR_45dB_16st_1.eps} 
%\caption{BER Vs. SNDR over Channel-A and TSTNR 40 dB.}
%\label{BERPQNRB}
%\end{figure}

%\begin{figure*}[ht]
%\centering
%\begin{multicols}{2}
%    \includegraphics[width=2in]{PAPR_RX_V1.eps} 
%    \caption{caption here}
%    \includegraphics[width=2in]{PAPR_RX_V1_224Gsymsec.eps}
%    \caption{caption here}
%    \end{multicols}
    
%\begin{multicols}{2}
%    \includegraphics[width=2.2in]{RSDNRBER_TSTNR_45dB_16st_1.eps}
%    \caption{caption here}
 %   \includegraphics[width=2.2in]{RSDNRBER_TSTNR_45dB_16st_1.eps}
 %   \caption{caption here}
%\end{multicols}
%\end{figure*}

The resulting PAPR distributions at Channel-A  output and the BER curves are presented in Fig. \ref{PAPRDIST}(a) and Fig. \ref{BERPQNRB}(a) respectively, for TSTNR 40 dB. 
As was shown in Fig. \ref{THEOR_G_T}(a),  the maximal gain is achieved in $\gamma$ -15 dB. However, since for practical implementation the shaping was applied over $Q$-point constellation rather than infinite set of points, the optimal BER performance was achieved in $\gamma$ of -14 dB and -3.9 dB for shaped 8-PAM and shaped 4-PAM systems, respectively.
The SNDR at which BER $10^{-6}$ is reached, the PAPR at Channel-A output and the required ENOB are summarized in Table \ref{GAINSSUM}. It can be seen that the shaped 8-PAM and 4-PAM systems achieve, overall shaping gains of 8.55 dB and 4.05 dB, respectively, compared to uniform 4-PAM transmission with TE. These gains translates to ENOB gain of 1.43 bit and 0.68 bit, respectively. 
Comparing the SNDR gain to the theoretical SNDR gain indicates that the online shaping scheme suffers from loss of 1.98 dB.

\begin{table} [H]
\begin{center}
    \begin{tabularx}{0.912\linewidth}{|c |  c |c |c|}
    \hline
    System & PAPR $10^{-4}$  & SNDR $10^{-6}$ & ENOB 
    \\ \hline    
    8-PAM uniform + TE & 10.35 dB  & 23.85 dB & 4.9 bit           \\ \hline
    4-PAM uniform + TE & 10.13 dB & 20.02 dB & 4.23 bit       \\ \hline
    8-PAM shaped & 5.3 dB & 16.3 dB   & 2.8 bit \\ \hline
    4-PAM shaped & 6.45 dB & 19.65 dB   & 3.55 bit  \\ \hline
    \end{tabularx}
    \caption{Rate 1.8 bits/symbol and TSTNR 40 dB over Channel-A, PAPR at channel output, SNDR and ENOB summary.}
    \label{GAINSSUM}
\end{center}
\end{table}

%The PAPR distributions at Channel-A output and the relationships between BER and RPQNR, for both shaped and uniform systems, are presented in  Fig. \ref{PAPRDIST} and Fig. \ref{BERPQNRB}, respectively, for TSTNR 45 dB.
%In this working point, the optimal BER performance was achieved in peak constraint $\gamma$ of -14 dB and -3.9 dB for shaped 8-PAM and shaped 4-PAM systems, respectively. Also shown in Fig. \ref{BERPQNRB}, the performance of an un-coded uniform 8-PAM and uniform 4-PAM systems, with a Decision Feedback Equalizer (DFE) \cite{DFE} at the receiver.   
%Fig. \ref{PAPRDIST} indicates that the shaped 8-PAM and 4-PAM systems achieve PAPR gains of 4.83 dB and 3.68 dB, respectively, compared to uniform 4-PAM transmission and gains of 5.1 dB and 3.85 dB, respectively, compared to uniform 8-PAM transmission.
% since the theoretical SNDR gain is 5.7 dB whereas the practical SNDR gain is 3.72 dB.  

In the case of rate 1.8 bits/symbol over Channel-B and TSTNR 40 dB, the maximal gain was obtained when constraining the peak power, $\gamma$, to -17 dB. The resulting PAPR distributions and the BER curves are presented in Fig. \ref{PAPRDIST}(b) and Fig. \ref{BERPQNRB}(b), respectively. As before, the metrics of interest are summarized in Table \ref{GAINSSUM1}. It can be seen that the shaped 8-PAM and 4-PAM systems achieve overall shaping gains of 10.65 dB and 5.45 dB, respectively, compared to uniform 4-PAM transmission with TE. These gains translates to ENOB gain of 1.78 bit and 0.91 bit, respectively. 
 
\begin{table} [H]
\begin{center}
    \begin{tabularx}{0.912\linewidth}{|c |  c |c |c|}
    \hline
    System & PAPR $10^{-4}$  & SNDR $10^{-6}$ & ENOB 
    \\ \hline    
    8-PAM uniform + TE & 11 dB  & 28.3 dB & 5.75 bit           \\ \hline
    4-PAM uniform + TE & 10.95 dB & 24 dB & 5.03 bit       \\ \hline
    8-PAM shaped & 5.3 dB & 19 dB   & 3.25 bit \\ \hline
    4-PAM shaped & 6.4 dB & 23.1 dB   & 4.12 bit  \\ \hline
    \end{tabularx}
    \caption{Rate 1.8 bits/symbol and TSTNR 40 dB over Channel-B, PAPR at channel output, SNDR and ENOB summary.}
    \label{GAINSSUM1}
\end{center}
\end{table}
\section{Conclusion}
\label{conclusion}
%======================================================================

A novel online shaping technique for PAPR reduction at the output of high-speed wireline channels has presented. The technique is effective to reduce the large ADC dynamic range requirement and by that the required ENOB, such that an overall gain is achieved compared to uniform transmission with TE at the receiver.
Theoretical analysis which provides a LB on the SNDR and theoretical shaping gains has derived as well.   
%It was further shown that the shaping scheme shapes the transmitted PSD as high pass, without using any filter, with a slope that depends on the peak constraint. At low enough peak constraint, the transmitted PSD approximates the channel inversion spectrum. The technique is therefore also effective for ISI reduction. 
In data rate of 200 Gbps and 400 Gbps, an overall ENOB gains was demonstrated to be up to 1.43 bit and 1.78 bit, respectively, compared to a uniform 4-PAM transmission with TE at the receiver side.

%===============================================================================

\end{document}